\documentclass[12pt]{article}
\usepackage{times}
\usepackage{geometry}
\usepackage[utf8]{inputenc}
\usepackage{enumitem,amssymb}
\usepackage{multicol}
\usepackage{ragged2e}
\newlist{thematic}{itemize}{8}
\setlist[thematic]{label=$\square$}
\usepackage{pifont}

\usepackage{mathptmx} 
\usepackage{graphicx}
\usepackage{amssymb}
\usepackage{epstopdf}
\usepackage{enumitem}
\usepackage{fancyhdr}
\usepackage{amsmath}
\usepackage[font={footnotesize, it},labelfont={footnotesize, bf}]{caption}
\usepackage{subcaption}
\usepackage{epstopdf}
\usepackage{wrapfig}
\usepackage{hyperref}
\usepackage{natbib}

\begin{document}
\raggedright
\huge
Astro2020 Science White Paper \linebreak

Exocometary Science \linebreak
\normalsize

\noindent \textbf{Thematic Areas:} \hspace*{60pt} $\boxtimes$ Planetary Systems \hspace*{10pt} $\boxtimes$ Star and Planet Formation \hspace*{20pt}\linebreak
$\square$ Formation and Evolution of Compact Objects \hspace*{31pt} $\square$ Cosmology and Fundamental Physics \linebreak
  $\square$  Stars and Stellar Evolution \hspace*{1pt} $\square$ Resolved Stellar Populations and their Environments \hspace*{40pt} \linebreak
  $\square$    Galaxy Evolution   \hspace*{45pt} $\square$             Multi-Messenger Astronomy and Astrophysics \hspace*{65pt} \linebreak
  
\textbf{Principal Author:}

Name:	Luca Matr\`a
 \linebreak					
Institution:  Center for Astrophysics $|$ Harvard \& Smithsonian, Cambridge MA, USA
 \linebreak
Email: luca.matra@cfa.harvard.edu 
 \linebreak
Phone:  617-495-7085
 \linebreak
 \vspace{-2mm}

\textbf{Co-authors:}
  \linebreak
Quentin Kral, Kate Su, Alexis Brandeker, William Dent, Andras Gaspar, Grant Kennedy, Sebastian Marino, Karin \"Oberg, Aki Roberge, David Wilner, Paul Wilson, Mark Wyatt\\
\vspace{4mm}
\textbf{Co-signers:}
  \linebreak
Gianni Cataldi, Aya Higuchi, A. Meredith Hughes, Flavien Kiefer, Alain Lecavelier des Etangs, Wladimir Lyra, Brenda Matthews, Attila Mo\'or, Barry Welsh, Ben Zuckerman\\
\vspace{4mm}
\textbf{Abstract:}\\
\justifying
\frenchspacing
Evidence for exocomets, icy bodies in extrasolar planetary systems, has rapidly increased over the past decade. Volatiles are detected through the gas that exocomets release as they collide and grind down within their natal belts, or as they sublimate once scattered inwards to the regions closest to their host star. Most detections are in young, 10 to a few 100 Myr-old systems that are undergoing the final stages of terrestrial planet formation. This opens the exciting possibility to study exocomets at the epoch of volatile delivery to the inner regions of planetary systems. 
Detection of molecular and atomic gas in exocometary belts allows us to estimate molecular ice abundances and overall elemental abundances, enabling comparison with the Solar Nebula and Solar System comets. At the same time, observing star-grazing exocomets transiting in front of their star (for planetary systems viewed edge-on) and exozodiacal dust in the systems' innermost regions gives unique dynamical insights into the inward scattering process producing delivery to inner rocky planets.
The rapid advances of this budding subfield of exoplanetary science will continue in the short term with the upcoming \textit{JWST}, \textit{WFIRST} and \textit{PLATO} missions. In the longer term, the priority should be to explore the full composition of exocomets, including species crucial for delivery and later prebiotic synthesis. Doing so around an increasingly large population of exoplanetary systems is equally important, to enable comparative studies of young exocomets at the epoch of volatile delivery. We identify the proposed \textit{LUVOIR} and \textit{Origins} flagship missions as the most promising for a large-scale exploration of exocometary gas, a crucial component of the chemical heritage of young exo-Earths.
\thispagestyle{empty}

\pagebreak
\setcounter{page}{1}

\noindent \textbf{\large 1\hspace{3mm} Background and Motivation} \\
\vspace{-3.0mm}

Understanding the origin of life stands as one of the main drivers of interdisciplinary research that includes chemistry, biology and geology as well as astronomy. 
The development of prebiotic molecules requires not only particular physical conditions, but also the presence of simple feedstock volatile molecules like water, cyanides and sulfides \citep[e.g.][]{Patel2015}. 
Young and dry rocky planets may be lacking in these volatiles \citep{Albarede2009}, requiring external delivery from volatile-rich \textit{comets}.
By comet, we refer to the IAU definition of \textit{a body made of rock and ice, typically a few kilometres in diameter}, and consider all minor bodies that contain a significant amount of ice, be they in stable orbits within their outer reservoirs, or be they scattered inwards and evaporating as they approach the central star.
Young comets, in the Solar Nebula and its extrasolar counterparts, form from the coagulation of dust and freeze-out of major gas species onto their surface. In the first $\sim$10 Myr of this gas-rich environment, comets may accrete to form icy worlds as seen in the outer reaches of the Solar System, or end up in stable long-period orbits, forming rings and broader belts. These belts, such as our own Edgeworth-Kuiper belt, are broadly termed \textit{planetesimal belts} or \textit{debris disks}. 


These belts of \textit{exocomets} survive the dissipation of protoplanetary disks and are later observed throughout the lifetime of the stars. Exocometary belts significantly more massive than today's Kuiper belt are present around at least $\sim$20\% of nearby stars, producing observable dust and losing mass over time through collisional grinding (see White Paper by Gaspar et al.). The most massive and easily detectable belts are therefore the youngest, at ages of 10 to 100 Myr. Crucially, this is the point at which planetary systems undergo the violent era of planetary assembly, when the dynamical action of recently formed planets on exocomets is most likely to produce inward scattering \citep[e.g.][]{Morbidelli2012} and volatile delivery to the inner, rocky young Earth analogs.  

Assessing the potential for volatile delivery in planetary systems is the key driver for research investigating the dynamics and composition of exocomets. 
Studies of solids within belts of exocomets have been successfully conducted for the past three decades (see White Papers by Gaspar et al., Su et al.). Here, we highlight the \textit{exocometary} aspect, i.e. studies of the volatile composition and inward scattering of exocomets, which are most relevant to the issue of volatile delivery.
These studies have seen significant advances in the past decade, giving rise to the field of \textit{Exocometary Science}.
In this White Paper, we summarize these advances and outline the observational requirements for the field to continue to prosper in the next decade and beyond.

\ \\

\begin{figure}[h]
\vspace{-13mm}
    \includegraphics[width=1.0\textwidth]{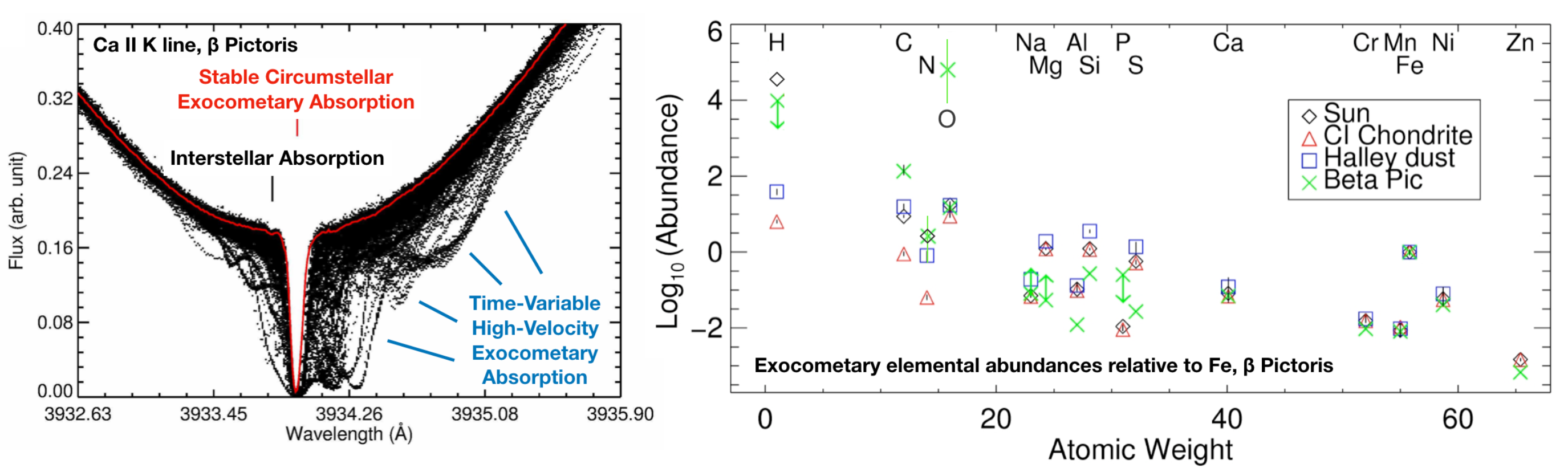}
  \vspace{-9mm}
  \caption{\textit{Left:} Ca II K line profile of $\beta$ Pictoris, showing a \textit{stable} absorption feature from exocometary gas in the outer belt, as well as time-variable, high-velocity components from star-grazing exocomets on eccentric orbits. Adapted from \citet{Kiefer2014}. \textit{Right:} Elemental abundances of exocometary gas in the outer disk of $\beta$ Pic, from studies of the stable absorption component. Adapted from \citet{Roberge2006}, with the addition of the oxygen abundance estimate of \citet{Brandeker2016}, and the hydrogen and nitrogen abundance from \citet{Wilson2017, Wilson2019}.}
  \vspace{-8mm}
\end{figure}

\newpage
\vspace{-5mm}
\noindent \textbf{\large 2\hspace{3mm}  The Emergence of Exocometary Science} \\

\vspace{-4mm}
The first evidence of exocomets dates back to 1975, when deep, sharp gas absorption components were observed superimposed to the Ca II H and K absorption lines of the nearby main-sequence star $\beta$ Pictoris \citep[][ also e.g. Fig. 1]{Slettebak1975}. More than 10 years later, a redshifted variable absorption component to this sharp line was discovered, and for the first time attributed to gas from \textit{falling evaporating bodies} (FEBs), i.e. star-grazing exocomets on highly eccentric orbits transiting at a few stellar radii in front of the star \citep{Ferlet1987, Beust1990}. 


\vspace{2mm}
\noindent \textbf{2.1\hspace{2mm} From discovery to \textit{Herschel}: the predominance of atomic gas} \\
In the last decade, 
evidence for exocomets mounted and exocometary science picked up significant momentum.
Extensive searches for cold CO gas at mm wavelengths beyond the original detection around 49 Ceti \citep{Zuckerman1995} were conducted, yielding new detections \citep[e.g.][]{Moor2011}. At the same time, \textit{Herschel} detected atomic oxygen (OI) and ionized carbon (CII) emission in a few systems \citep[e.g.][]{RiviereMarichalar2012, RiviereMarichalar2013}, 
and confirmed that the stable UV absorption lines seen in $\beta$ Pic arise from a gas disk in Keplerian rotation around the central star \citep{Brandeker2004,Cataldi2014}. 
\textit{Herschel} and later sub-millimeter observations indicate that atoms dominate the gas mass over molecules \citep{Kral2016a, Higuchi2017}, in contrast to younger protoplanetary disks. This was the first indication of tenuous gas disks that are optically thin to photodissociating UV radiation.

\vspace{2mm}
\noindent \textbf{2.2\hspace{2mm} The exocometary origin of CO gas, and extracting exocometary ice compositions} \\
ALMA confirmed the existence of tenuous gas disks by detecting 6 systems hosting very low masses of CO gas, around $\sim$10$^{-7}$ to 10$^{-5}$ M$_{\oplus}$, co-located with the planetesimal belts \citep[e.g.][]{Dent2014, Marino2016, Marino2017a, Matra2017b, Matra2019a, Booth2019}. Even if unreasonably large amounts of unseen H$_2$ were present, the CO would be optically thin to photodissociating UV light, being destroyed on timescales of order $\sim$100 years. This implies that for us to see the CO, it must have been recently produced, which is evidence for continuous release from exocometary ices.

\begin{wrapfigure}{r}{0.55\textwidth}
\vspace{-11mm}
    \includegraphics[width=0.55\textwidth]{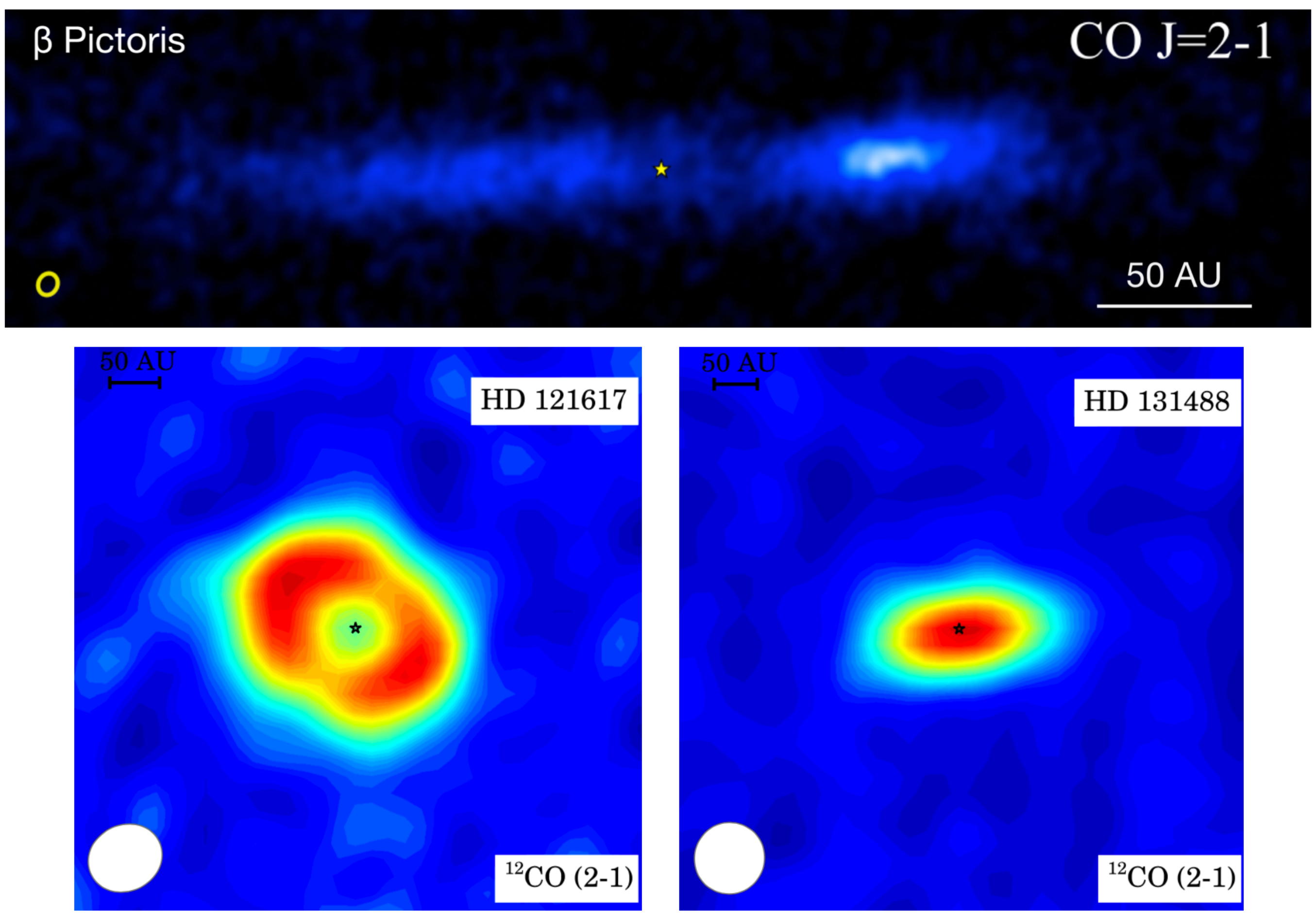}
  \vspace{-7mm}
  \caption{Exocometary CO within the $\beta$ Pic belt \citep[top, adapted from][]{Matra2017a}, and two young A-type stars in Scorpius-Centaurus \citep[bottom, adapted from][]{Moor2017}. The CO asymmetry around $\beta$ Pic indicates preferential locations of molecular gas production followed by photodissociation in less than an orbital timescale. }
  \vspace{-7mm}
\end{wrapfigure}

Gas (including CO and/or atomic species) has been detected in 14 more systems \citep[e.g.][]{Lieman-Sifry2016}, but its origin in these systems remains unconfirmed, as enough gas mass could be present to prolong the photodissociation lifetime \citep[e.g.][]{Kospal2013}, although this does not rule out an exocometary origin \citep{Kral2018}. 
One key property of exocometary gas is that it is expected to be H$_2$-poor, in contrast with protoplanetary disks and ISM gas. There is some evidence in favour of this, including  
CO line ratios showing low excitation temperatures and small vertical scale heights measured for nearly edge-on disks \citep{Matra2017a, Hughes2017}. 
Most gas-bearing belts are found around A-type stars \citep[Fig. 2, also][]{Moor2017}, with a high CO detection rate (69$^{+9}_{-13}$\%) compared to later-type stars (7$^{+13}_{-2}$\%), where this trend can be explained by exocometary release combined with observational detection bias \citep{Matra2019a}. The current occurrence rates are consistent with all exocometary belts releasing gas at some level, with all detections and upper limits compatible with compositions observed for Solar System comets \citep{Kral2017, Kral2018,Matra2019a}.


Detecting molecular gas allows us to probe the ice content of exocomets through a simple steady state argument, assuming that the observed gas species are released from the ice within the same collisional cascade that produces the observed dust \citep{Zuckerman2012, Matra2015, Matra2017b, Kral2016a}. 
The CO(+CO$_2$) ice mass fractions measured in 6 exocometary belts are within an order of magnitude of Solar System comets \citep[e.g.][]{Matra2017b}, confirming that detection of exocometary gas from Solar System-like exocomets has become feasible, and hinting at a similarity in cometary compositions across planetary systems.

Other species commonly seen in Solar System comets should also be released, as indicated by the presence of atomic N gas \citep{Wilson2019}, a low C/O ratio in the gas favouring oxygen around $\beta$ Pictoris \citep[which could point to water and/or CO$_2$ release,][]{Brandeker2016, Kral2016a}, and models predicting efficient outgassing through UV photodesorption or high-velocity collisions of the smallest grains \citep{Grigorieva2007, Czechowski2007}.
However, molecular species other than CO have much shorter photodissociation timescales, making their detection \citep[aside from HCN and CN, which should be potentially detectable with ALMA,][]{Matra2018a} challenging with currently available facilities. 

\vspace{2mm}
\noindent \textbf{2.3\hspace{2mm} Exocometary atomic gas: viscous evolution and bulk composition} \\
The atomic gas produced by molecular photodissociation constitutes the bulk of the exocometary gas mass, and in about a viscous timescale, will spread radially to produce a steady state accretion disk \citep{Kral2016a}. 
The $\alpha$ viscosity could be high if the ionization fractions are large and the magneto-rotational instability (MRI) is at play \citep{Kral2016b}. So far, however, low $\alpha$ values of order 10$^{-3}$-10$^{-4}$ (or a recent onset of gas release) are required to explain the inner holes resolved by atomic carbon (CI) observations \citep{Cataldi2018, Kral2018}.

Observing atomic gas allows us to study elemental abundances, providing another route to exocometary compositions. Given the current lack of far-IR space facilities, UV absorption studies of a few edge-on belts are best suited for elemental inventories \citep[e.g. Fig. 1 right,][]{Roberge2006}, and can be used to compare with Solar and Solar System cometary and asteroidal abundances. $\beta$ Pictoris is the only star with a full inventory, and shows sub-Solar H and super-Solar C and O abundances. While the sub-Solar H content \citep{Wilson2017} further confirms the exocometary origin of the gas, the super-Solar C (in its ionized state) acts to stabilize the gas counteracting outward radiation pressure that affects metallic species \citep{Fernandez2006, Xie2013}, and explaining their observed spatial distribution \citep{Brandeker2004}.

\vspace{2mm}
\noindent \textbf{2.4\hspace{2mm} Evidence for inward scattering} \\
Inward scattering is fundamental to connect the outer exocomet population discussed above to the potential for volatile delivery to the inner planets.
The first extrasolar evidence for inward scattering is the presence of time-variable high-velocity narrow gas absorption lines in systems viewed nearly edge-on \citep[e.g. Fig. 1, left, also][]{Ferlet1987}. These are produced by star-grazing exocomets on eccentric orbits evaporating as they approach their star \citep[at a few tens of stellar radii, e.g.][]{Beust1990}. At these locations, temperatures are high enough \citep[e.g.][]{Vidal-Madjar2017} for refractory elements to evaporate together with volatiles, producing the metallic lines observed. Around $\beta$ Pictoris, 8 years of monitoring revealed two populations of absorption lines with starkly different depths and widths, attributed to two families of exocomets at different distances with different outgassing rates \citep{Kiefer2014}, and accelerating on their way to the star \citep{Kennedy2018b}.
Evidence of star-grazing exocomets is also present around $\sim$20 other stars hosting outer planetesimal belts \citep[e.g.][]{Welsh2018}, and in $\sim$78-88\% of edge-on systems with evidence for cold gas at larger radii \citep{Rebollido2018}.

In addition to variable gas absorption lines, broadband photometric variability in the form of asymmetric transits has long been predicted as a sign of exocometary activity \citep{LecavelierdesEtangs1999}, and finally confirmed in the past year with the discovery of asymmetric transits in \textit{Kepler} light curves \citep{Rappaport2018, Kennedy2019}, as well as potentially in the dimming events of the star KIC8462852 \citep{Wyatt2018b}. 
Finally, the presence of exozodiacal dust with temperature from hundreds to thousands of K in the innermost regions of planetary systems (see White Paper by Mennesson et al.) may also be produced by inward scattering. For example, in the Gyr-old $\eta$ Corvi system, host to both exozodiacal dust as well as an outer planetesimal belt, detection of faint CO emission in the inner regions could be evidence for sublimation of snowline-crossing exocomets \citep{Marino2017a, Marino2018a}. \\

%





\vspace{-2mm}
\noindent \textbf{\large 3\hspace{3mm} Outlook and Recommendations} \\
\vspace{-4mm}

Following the successes of this past decade, exocometary science will expand in two main directions.
First, studies of exocometary dust in the inner regions of planetary systems, from exozodis to transit observations, will inform models of inward scattering and provide rates of inward delivery across planetary systems. This will be achieved with upcoming high-contrast imaging observatories such as \textit{JWST} and \textit{WFIRST} in the next decade (and proposed missions like \textit{HabEx} and \textit{LUVOIR} in the longer term). At the same time, high-accuracy photometric missions like \textit{TESS} and \textit{PLATO} will inform us on how common transiting exocomets are by detecting new transits toward the shallow end of the transit depth distribution \citep{Kennedy2019}.

Second, we stress the importance of gas observations to access exocometary compositions.
The 2020s will see a continued effort with existing facilities like \textit{ALMA} and \textit{HST} and new opportunities with the upcoming \textit{JWST}, whose spectral coverage will include rovibrational transitions of molecules such as CO$_2$, H$_2$, H$_2$O and OH. However, detection of emission lines could be hampered by the fact that exocometary gas currently observed within exocometary belts, at tens of au from the central star, is cold (at most a few tens of K). Then, excitation and hence detectability of these high energy transitions with \textit{JWST} will rely on UV and IR pumping (i.e. radiative absorption to electronic and rovibrational levels producing fluorescence) from the central star, rather than collisions \citep[as described in e.g.][]{Matra2018a}, which will require detailed excitation modelling. At the same time, large surveys will not be possible due to \textit{JWST}'s large slewing overheads.


For the longer term, the main priorities should be 1) to fully explore the chemical composition of exocomets during the volatile delivery epoch, and 2) to significantly increase the number of planetary systems with detected compositional signatures, enabling population studies. 
Proposed radio interferometers like the next generation Very Large Array (ngVLA) and the Square Kilometer Array (SKA, phase 1) would largely improve on the current VLA capabilities for observations of exocometary H, NH$_3$, and OH. However, initial sensitivity considerations indicate that detection of these molecules will be at best sufficient to detect the most gas-rich systems.

As mentioned earlier, atoms (both neutral and ionized species) are expected to dominate the gas abundance and can be used to obtain inventories of elemental abundances (e.g. Fig. 1). This is best done at UV wavelengths, where the strongest transitions of dominant species (H, C, N and O) are located. The proposed \textit{LUVOIR} flagship mission with its LUMOS instrument promises a considerable sensitivity and spectro-spatial resolution improvement compared to \textit{HST}. In addition to absorption studies in edge-on systems, LUMOS will have the capability to provide spatially resolved maps of emission for the same dominant atomic species. Since multiple ionization stages and transitions can be detected in the UV, LUMOS could also provide maps of environmental conditions, as well as elemental abundances \citep[e.g.][]{Nilsson2012}. This would make LUMOS an extremely powerful instrument for detailed characterization of exocometary atomic gas.

\begin{wrapfigure}{r}{0.38\textwidth}
\vspace{-6mm}
    \includegraphics[width=0.38\textwidth]{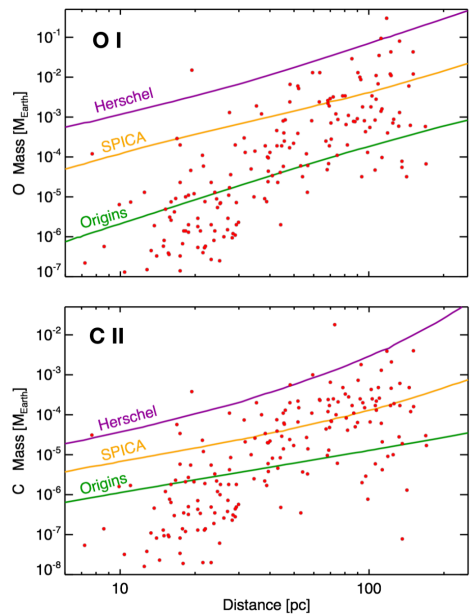}
  \vspace{-8mm}
  \caption{Exocometary O and C gas masses predicted by the steady state release model (red points) in planetary systems with known exocometary belts. Lines represent mass sensitivities of different far-IR facilities, showing how the sensitivity of \textit{Origins} would revolutionize the field by yielding $>100$ new detections, enabling population studies. Adapted from \citet{Kral2017}.}
  \vspace{-5mm}
\end{wrapfigure}

The far-IR, on the other hand, offers the best window into observing water. This is because cold H$_2$O and its longer-lived photodissociation product OH have their lowest-energy levels, and hence strongest transitions, in the far-infrared (e.g., 119 $\mu$m for OH). Ongoing work shows that the OSS instrument on the proposed \textit{Origins} flagship mission, with its factor $\sim$1000 sensitivity improvement over \textit{Herschel}, will easily detect OH produced by water photodissociation in nearby planetesimal belts \citep{Matra2018a}. Water is the most abundant species in Solar System comets and plays a key role in the development of life; we therefore stress the importance of proving its existence on exocomets during the epoch of volatile delivery.

With \textit{Origins}, detection of dominant atomic species C and O through their low-energy, fine structure transitions is expected to lead to new exocometary gas detections around $\gtrsim$100 targets \citep[Fig. 3, also][]{Kral2017}. Furthermore, many nearby exocometary belts will be spatially and spectrally resolved, allowing us to identify the gas production location and study its dynamical evolution. Spectrally resolving the lines (with the proposed R$\sim$40000 for \textit{Origins}) is also crucial for detecting faint systems, as it allows for more favorable line-to-continuum ratios. Therefore, \textit{Origins} promises to be a unique tool both for detection of water during volatile delivery as well as for sensitive, unbiased surveys to newly detect gas and access compositions in a large number of extrasolar planetary systems. This can only partially be achieved with other far-IR instruments under consideration, such as \textit{SOFIA}/HIRMES and \textit{SPICA}/SAFARI, which would bring a more modest sensitivity and/or resolution improvement compared to \textit{Origins}.


Thanks to recent sensitivity advances, exocometary science has seen significant expansion in the last decade, establishing itself as an emerging component of exoplanetary astronomy. Studies of exocomets take advantage of multiple techniques and probe the composition and dynamics of emerging planetary systems, providing unique insights into volatile delivery. The field will continue to expand in the 2020s thanks to current and upcoming facilities such as \textit{JWST}, \textit{WFIRST} and \textit{PLATO}. To truly revolutionize exocometary science in the 2030s, we highlight the need of either a large UV mission such as \textit{LUVOIR} to study and map elemental abundances of the main, atomic exocometary gas carriers, or a far-infrared surveyor mission such as \textit{Origins} to study water at the epoch of volatile delivery and survey exocometary gas in a large number of planetary systems.






\ \\
\begin{multicols}{2}
\bibliographystyle{apj}
\bibliography{lib}
\end{multicols}

\end{document}